\documentclass[sn-basic,Numbered]{sn-jnl}

\usepackage{graphicx}%
\usepackage{multirow}%
\usepackage{amsmath,amssymb,amsfonts}%
\usepackage{amsthm}%
\usepackage{mathrsfs}%
\usepackage[title]{appendix}%
\usepackage{xcolor}%
\usepackage{textcomp}%
\usepackage{manyfoot}%
\usepackage{booktabs}%
\usepackage{algorithm}%
\usepackage{algorithmicx}%
\usepackage{algpseudocode}%
\usepackage{listings}%

\theoremstyle{thmstyleone}%
\theoremstyle{thmstyletwo}%

\theoremstyle{thmstylethree}%

\begin{document}

\title[\texttt{penalizedclr}: an R package for penalized conditional logistic regression for integration of multiple omics layers]{\texttt{penalizedclr}: an R package for penalized conditional logistic regression for integration of multiple omics layers}


\author*[1,2]{\fnm{Vera} \sur{Djordjilovi\'c}}\email{vera.djordjilovic@unive.it}

\author[2]{\fnm{Erica} \sur{Ponzi}}

\author[3,4]{\fnm{Therese}\sur{Haugdahl Nøst}}

\author[2]{\fnm{Magne} \sur{Thoresen}}

\affil[1]{\orgdiv{Department of Economics}, \orgname{Ca' Foscari University of Venice}, \orgaddress{\city{Venice},  \country{Italy}}}

\affil[2]{\orgdiv{Department of Biostatistics}, \orgname{University of Oslo}, \orgaddress{\city{Oslo},  \country{Norway}}}

\affil[3]{\orgdiv{Department of Public Health and Nursing}, \orgname{Norwegian University of Science and Technology}, \orgaddress{\city{Trondheim}, \country{Norway}}}

\affil[4]{\orgdiv{Department of Community Medicine, Faculty of Health Sciences}, \orgname{The Arctic University of Norway}, \orgaddress{ \city{Troms\o}, \country{Norway}}}



\abstract{\textbf{Background:} The  matched case-control design, up until recently mostly pertinent to epidemiological studies, is becoming customary in biomedical applications  as well. For instance, in omics studies, it is quite  common to compare cancer and healthy tissue from the same patient. Furthermore, researchers today routinely collect data from various and  variable sources that they wish to relate to the case-control status.  This highlights the need to develop and implement   statistical methods that can take these tendencies into account.
\textbf{Results:} We  present an R package \texttt{penalizedclr}, that provides an implementation of the penalized conditional logistic regression model for analyzing matched case-control studies. It allows for different penalties for different blocks of covariates, and it is therefore particularly useful in the presence of multi-source omics data. Both L1 and L2 penalties are implemented. 
Additionally, the package implements stability selection for variable selection in the considered regression model.
\textbf{Conclusions:} The proposed method fills a gap in the available software for  fitting high-dimensional conditional  logistic regression model accounting for the matched  design and block structure of   predictors/features. The output consists of a set of selected variables that are significantly associated with case-control status. These features can then be investigated in terms of functional interpretation or validation in further, more targeted studies. 
}




\maketitle

\section{Background}\label{sec1}

The matched case-control design is widely employed in biomedical studies, since matching on potentially confounding variables can significantly improve efficiency and statistical power, while mitigating the effect of potential confounders. This design has become popular in studies involving high-throughput assays, leading researchers to propose novel methods for the analysis of high-dimensional matched data, also with the aim of feature or variable selection  \citep{liang2018review}.  As many of these ignore the study design and apply methods not designed for the matched design, this strategy can lead to sub-optimal results \citep[see for instance][]{balasubramanian2014variable, shomal2020matched}  and potentially missing some important associations.   
A classical method for taking into account the matched design is offered by conditional logistic regression, either applied to each variable individually or applied to all variables jointly in a multivariate model  \citep[see for instance][]{ balasubramanian2014variable} which is the approach we consider here.

Studies containing several types of high-dimensional measurements for each individual -- for instance, DNA methylation, copy number variation and mRNA expression -- are becoming increasingly common.  Integrating such heterogeneous data layers poses an additional challenge to variable selection, as the optimal penalty parameters can vary across different layers.  An intuitively simple solution is to generalize a well-investigated method of penalized conditional logistic regression to allow for different penalties for different data layers. This approach can be particularly useful when the proportions of relevant variables are expected to vary across layers. 

The method proposed here is similar in spirit to the popular IPF-lasso \citep{boulesteix2017ipf} and IPFStructPenalty \citep{ISI:000530548200002}. The main difference  is the overall aim:  the two cited  methods are geared towards prediction in generalized linear models, we focus on variable selection instead. This choice is not a matter of preference: in  conditional logistic regression models, intercept terms are treated as nuisance, rendering predictions for new observations impossible. In view of this, the proposed method is designed  to address the initial challenge of selecting promising biomarker candidates. 

Results of variable selection procedures in high dimensional settings are known to suffer from limited replicability. To address this issue, our package provides an implementation of stability selection, a general method in which results of the selection procedure are aggregated over  different data subsamples \citep{meinshausen2010stability}. To develop good prediction algorithms useful from a diagnostic and clinical  perspective, a biological  interpretation of the selected candidates would be conducted and they should be  further investigated in a prospective study.

\section{Implementation}\label{implementation}

\texttt{penalizedclr} is implemented in R and available from CRAN. A development version is also available from github \url{https://github.com/veradjordjilovic/penalizedclr}. 

In what follows, we describe the two main functions of the  package,  \texttt{penalized.clr}, estimating a penalized \textbf{c}onditional \textbf{l}ogistic \textbf{r}egression model allowing for different penalties for different blocks of covariates, and \texttt{stable.clr.g} performing stability selection of variables in the penalized conditional regression model. We then discuss other important aspects of the implementation, such as the choice of the penalization  parameters and computation time.

\subsection{\texttt{penalized.clr} function}
This is a wrapper function for the \texttt{penalized}  function of the well-established  \texttt{R} package of the same name \citep{goeman2010l1,penalized}. A routine for conditional logistic regression is not directly available in \texttt{penalized}, but we exploit the fact that the likelihood of a conditional logistic regression model is the same as that of a Cox model with a specific data structure. In the input, we need to specify the response vector, the stratum membership of each observation, i.e. in  case of 1:1 matching, the id of the  case-control pair the  observation belongs to;  the overall matrix of covariates to be penalized, the sizes of the blocks of covariates and the (L1) penalties to be applied to each block.   The output is a list including the estimated regression coefficients, along with other useful information regarding the fitted model. It should be stressed, that the vector of penalties has no default value and thus needs to be specified by the user. 

\subsection{\texttt{stable.clr.g} function}
\label{stability}To increase the replicability of research findings -- in this case selected variables -- we aim to select
  variables that are robust to small perturbations in the data.  To this end, we have implemented  stability selection \citep{meinshausen2010stability} in the function \texttt{stable.clr.g}. Here, most of the required input arguments are the same as in \texttt{penalized.clr},  with the argument \texttt{lambda.list} replacing  \texttt{lambda}. 
The argument  \texttt{lambda.list}  consists of vectors of $L_1$ penalties to be applied to each penalized block of covariates. Each vector has length equal to the number of blocks. For each vector, $2B$ random subsamples of  $\lfloor n/2 \rfloor$ (out of the total of $n$) matched pairs  are taken and a penalized model is estimated ($B = 100$ by default).  The factor $2$ in $2B$ is due to a variant of stability selection that includes complementary pairs of subsamples \citep{shah2013variable}. 
For each variable and vector of penalties, a  selection probability is estimated as the proportion of fitted models in which the associated coefficient estimate is different from zero. Finally, the estimate of the selection probability of a variable is obtained by taking the maximum selection probability over all considered penalty vectors.  The user can then select the variables whose estimated selection probability is above a desired threshold, typically in the range $0.55 - 0.9$.

\subsection{Data adaptive choice of penalty parameters}
\label{data_adaptive}
The user needs to specify penalties to be applied in the main functions. In general, choosing the appropriate amount of penalization is challenging, and even more so in the presence of multiple blocks of predictors with different penalties. The problem can be decomposed into two subproblems to be solved independently, as follows. Let $\boldsymbol{\lambda} = (\lambda_1, \lambda_2, \ldots, \lambda_P)$ represent a vector of $L_1$ penalties, where $\lambda_i$ is the penalty applied to the $i-$
th block, and $P$ is the number of blocks. Then we can write $\boldsymbol{\lambda} = \lambda_1(1, \lambda_2/\lambda_1, \ldots, \lambda_P/\lambda_1)$, where $\lambda_1$ can be viewed as the overall level of penalization, while the vector $(1, \lambda_2/\lambda_1,  \ldots, \lambda_P/\lambda_1)$ represents the relative penalties with respect to the first block. In \texttt{ipflasso}, this vector is referred to as the vector of penalty factors. Our package offers two functions: \texttt{default.pf} that searches for the data adaptive vector of penalty factors (see below), and \texttt{find.default.lambda} that for a given vector of penalty factors  finds  $\lambda_1$ that minimizes the cross-validated log-likelihood deviance \citep{reid2014regularization}. 

To find a data adaptive vector of penalty factors, we follow the heuristic approach of \cite{schulze2017}. In this extension of the original IPF-lasso method, a tentative conditional logistic regression model is fitted to all covariates, and for each block, the (relative) penalty is set to be inversely proportional to the mean of the estimated coefficients pertaining to that block. In this way, a block with larger estimated  coefficients will have a lower penalty, and vice-versa. This step can be performed for each block separately, i.e. by fitting $P$ tentative models, or jointly with all blocks included within a single model, see argument \texttt{type.step1}. Once a vector of penalty factors is obtained in this way, we can call \texttt{find.default.lambda} to find the value of $\lambda_1$ determining the overall extent of penalization. For more details, we refer to \cite{schulze2017} and the \texttt{penalizedclr} package documentation. 

\subsection{Elastic net penalty}
 The main focus of the package is on $L_1$ or lasso penalty which, resulting in sparse estimated models, is appropriate for  variable selection. Nevertheless, it is well-known that with $L_1$ penalty,  the presence of highly correlated variables 
 can have a negative impact on selection stability \citep{kirk2013balancing}. Adding a small $L_2$ or ridge penalty can alleviate this issue: our implementation offers this possibility by including the mixing parameter \texttt{alpha}, see package documentation for details. 

 \subsection{Computation time}
 The computational cost of estimating a penalized conditional logistic model with a given vector of penalties  equals the cost of estimating a penalized Cox model.  The time consuming part of the analysis is stability selection, which requires fitting $2Bs$ models, where $s$ is the number of the vectors of penalties in \texttt{lambda.list}.  Fortunately, stability selection is highly amenable to parallelization, which greatly  reduces computation time  especially when using a cluster of computers (see argument \texttt{parallel} of function \texttt{stable.clr.g}).

\section{Results}\label{sec2}

\subsection{Simulation study}
We illustrate the proposed method with a small simulation study. This simulation study is by no means meant to be exhaustive since many different simulation settings can be envisioned. The main  purpose of this study  is to illustrate some of the numerous factors that influence performance of the proposed method in  real applications.
The R code files for reproducing the results reported here  are available on github \url{https://github.com/veradjordjilovic/Simulations_penalizedclr}.

We considered six different settings described in Table \ref{tab1}, where $p_i$ and $a_i$ denote the dimension and the number of active variables in block $i$, respectively, while $\beta_i$ is the coefficient of an active variable in block $i$, $i=1,2$. Common for all settings is the number of blocks (2), the number of matched pairs ($200$), the total number of covariates ($100$) and the total number of active variables ($20$). 

For each setting, we generated 100 datasets, to which we applied a variable selection procedure based on conditional logistic regression as follows.  First, we computed data adaptive penalties, as described in Section \ref{data_adaptive}.  Next, we ran stability selection with $B=50$ on penalized conditional logistic regression with these penalties (Section \ref{stability}). Finally, covariates with selection probability exceeding $0.55$ were selected. 

We evaluated performance by measuring power, defined as the proportion of active variables identified by our procedure, and false discovery rate (FDR), defined as the proportion of false discoveries among all discoveries; in this case, the proportion of inactive variables among the selected variables. Power and FDR were averaged over 100 datasets. 

Results are shown in Table \ref{tab1}. We see that the power is lowest in settings 1, 4 and 6, in which either there is no (considerable) difference in the proportion of active variables in the two blocks (1 and 6) or the signal in one of the blocks is relatively weak. On the other hand, the highest power is achieved in setting 3, in which all active variables belong to the first block. Good power can also be observed in setting 5, where the majority of active  variables is in the first block. As for the empirical FDR, it seems comparable across settings, varying in the range 0.18 - 0.26. 

\begin{figure}
\includegraphics[width = 0.99\textwidth]{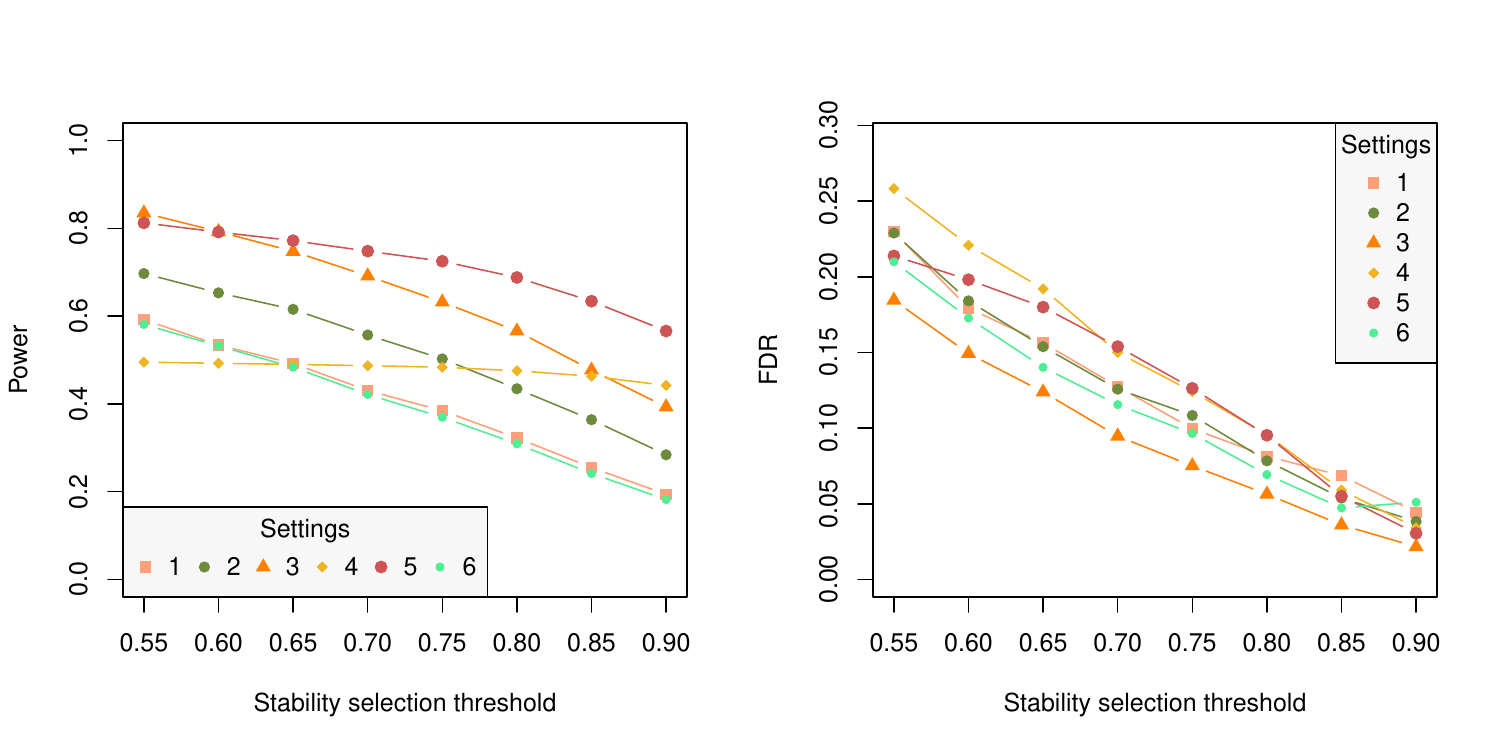}
\caption{\label{fig1} Empirical power and false discovery rate as a function of the threshold of the proposed variable selection procedure in 6 considered settings.}
\end{figure}

We set the threshold for selection to 0.55, which is at the low end of the suggested range (Section \ref{stability}).  To evaluate the impact of this choice, we computed the empirical power and FDR  for a grid of potential thresholds across the suggested range $(0.55 - 0.9)$. Results are shown in Figure \ref{fig1}. 

As expected, both power and FDR decrease with an increasing threshold for selection, since a stricter criterion for selection leads to  fewer selected variables, both active and inactive. Ordering of the settings is largely preserved across different thresholds (with some exceptions, for instance, the power for settings 3 and 5). Interestingly, setting 4 stands out from the rest: while its FDR decreases with the increasing threshold, as expected, its power remains constant. Recall that in setting 4, the number of active variables is equal among the two blocks, but the  signal strength  in  the second block is lower. Indeed, this signal seems to be too low to be picked up, and the variable selection procedure selects  only the variables of the first block.

\begin{table}
\caption{\label{tab1} Simulation study: description of simulation settings and the related performance. }
\begin{tabular}{l|llllll|ll}
\hline
Setting & \multicolumn{6}{c}{Parameters} & \multicolumn {2}{c}{Performance}\\
 & $p_1$ & $p_2$ & $a_1$ & $a_2$ & $\beta_1$ & $\beta_2$ & Power & FDR\\
\hline
1 & 50 & 50 & 10 & 10 & 4 & 4 & 0.59 & 0.23 \\
2 & 50 & 50 & 3 & 17 & 4 & 4 & 0.70 & 0.23\\
3 & 50 & 50 & 20 & 0 & 4 & 0 & 0.84 &  0.18\\
4 & 20 & 80 & 10 & 10 & 4 & 1  & 0.50 & 0.26
\\
5 & 20 & 80 & 15 & 5 & 4 & 4 & 0.81 & 0.21 \\
6 & 20 & 80 & 5 & 15 & 4 & 4  & 0.58 & 0.21\\
\hline

\end{tabular}
\end{table}

The main purpose of the presented simulation study  and the data application is to illustrate the possibilities and limitations of the proposed method. No comparison with other methods was reported, since, to the best of our knowledge, there are no other methods that implement  penalized  estimation of the conditional logistic regression model with multiple blocks of predictors. 

The small simulation study has showed, in line with the reported  results  for the IPF-lasso,  that taking  into account the block structure of predictors brings an advantage when the blocks are  indeed different, in terms  of signal strength and/or the number  or  proportion  of active variables. Otherwise, it is of course beneficial to treat all variables on equal standing, since in that case  we  are  dealing with  fewer tuning parameters.

\subsection{Application: multiomics analysis in lung cancer} \label{data_app}
To illustrate the proposed method in  practice, we consider a lung cancer matched case-control study nested within the Norwegian Women and Cancer Study (NOWAC) \citep{lund2008cohort}, a prospective cohort study. Our data consist of 125 case-control pairs matched by time since blood sampling and year of birth, identified in the
NOWAC cohort. Methylation levels and gene expression were measured in peripheral blood. 
We have focused on CpGs and genes that have previously been reported to be associated with smoking. In particular, we considered a list of CpGs differentially methylated between current smokers and nonsmokers according to \cite{joehanes2016epigenetic}. Since the total number of reported CpGs, 18760, precludes us from including them all in a considered multivariate model, we have selected the top 5000 CpGs according to their reported \(p\)-values. After restricting attention to complete observations, we were left with 4370 CpGs. Similarly, we considered a list of differentially expressed genes between current smokers and nonsmokers reported in \cite{huan2016whole}. Here, of the 1270 reported genes, in NOWAC we have information on 943 which we included in our analysis. 

Our goal was to select genes and CpGs that are associated with lung cancer status. Assuming a conditional logistic regression model, this amounts to selecting variables in the joint model that have a non-zero coefficient. 

We started our analysis by searching for the data adaptive vector of penalty factors. We  set $\alpha=0.6$ and ran the function \texttt{default.pf} three times, since this function relies on cross validation for selecting the penalty in the tentative model producing results that may vary between runs. In our case, the average vector of penalty factors was  proportional to $(1, 3.6)$. We then ran \texttt{find.default.lambda}, to find $\lambda_1 = 5.3$. 

For stability selection, we have considered the following list of penalty vectors: $(5,1), (5,2), (5,5), (5,10), (5,15), (5,20)$. We intentionally included combinations of penalties that appear to be far from the estimated data adaptive penalty factor, both to allow for less overall penalization and to explore different relative penalties for the two blocks. Our motivation comes from the observation that when conducting stability selection, it is more desirable to err on the side of too little penalization than too much.  In the former case, non-active variables are expected to vary randomly across different subsamples and achieve low selection probability. In the latter case, however, the large amount of penalization might negatively affect the power to identify active variables.

We set $0.55$ as the threshold for selection, and ended up with selecting two CpGs:  cg27039118 (estimated  selection probability: $0.56$), cg17065712 ($0.56$), and four genes: \textit{MAPRE2} ($0.63$), \textit{KCNMB1} ($0.78$), \textit{ATP1B1} ($0.61$)  and
\textit{SLC9A2} ($0.6$). Although they were all included in the analysis based on their reported association with smoking, none of these selected genes nor CpGs seem to have an established link to lung cancer.

\section{Conclusions}\label{conclusions}

In this work we have presented our implementation of the algorithm that allows for  fitting high dimensional conditional logistic regression model with covariates coming from different data sources.  The output of the proposed method is a set of  variables significantly associated with case-control status. To the best of our knowledge,  no such software has so far  been available  in the statistical software R. 

In the simulation study and the data application, we  considered 1:1 matching, but the proposed method is suitable  also  for  a general  1:$k$ matching, for $k\geq 1$, where each  case is matched to $k$ controls.

In our implementation, we have  opted for a data adaptive method for selecting penalty parameters that estimates tentative penalized model(s) and assigns   less penalty to blocks that have higher mean estimated coefficients.  Of course, there are many other sensible options for the choice of data adaptive penalty factors (see \texttt{ipflasso} R package). The user is free to combine the  proposed estimation procedure with an  arbitrary procedure for selecting  penalty parameters.

We have implemented stability selection with the aim of  stabilizing  the  obtained results in terms of selected  variables. However, 
stability  selection  can also be used  for Type  
1 error control. In particular, \cite{meinshausen2010stability} show  how  to bound the expected number of selected inactive  variables by means of stability  selection. Nevertheless,  their method  for  ensuring error control  relies on a  nontrivial  choice of tuning parameters,  which is an interesting research question on its own.  For this reason, we did not pursue this question in the present contribution.

\section{Availability and requirements
}
\texttt{penalizedclr} is implemented in R. Release versions are available on CRAN and work on all major operating systems. The development version is available at \url{https://github.com/veradjordjilovic/penalizedclr}.\\
\textbf{Project name:} \texttt{penalizedclr} R package\\
\textbf{Project home page:} \url{https://CRAN.R-project.org/package=penalizedclr}\\
\textbf{Operating system(s):} Platform independent.\\
\textbf{Programming language:}  R\\
\textbf{Other requirements:} No.\\
\textbf{License:} MIT + file LICENSE\\
\textbf{Any restrictions to use by non-academics:} No.

\section{Declarations}

\textbf{Ethics approval and consent to participate
.}
All participants gave written informed consent and the study was approved by the Regional Committee for Medical and Health Research Ethics and the Norwegian Data Inspectorate. More information is available in \cite{lund2008cohort}.\\
\textbf{Consent for publication.} Not applicable.\\
\textbf {Availability of data and materials.} Data analyzed in Section \ref{data_app} cannot be shared publicly because of local and national ethical and security policies. Data access for researchers will be conditional on adherence to both the data access procedures of the NOWAC study and the UiT, The Arctic University of Norway (contact: Tonje Braaten tonje.braaten@uit.no) in addition to approval from the local ethical committee.\\
\textbf{Competing interests.} The authors declare that they have no competing interests.\\
\textbf {Funding.} This research has been funded by grant no. 248804 and 262111 of the Norwegian Research Council.\\
\textbf{Authors' contributions.}
VD and MT conceived the research idea. VD, EP and THN conducted the statistical analyses. THN was responsible for the acquisition of data and the biological interpretation of the results. VD wrote the manuscript, with inputs from all authors. All authors gave final approval.

\bibliography{sn-bibliography}

\end{document}